\documentclass[12pt,preprint,showpacs,preprintnumbers]{revtex4}
\usepackage{amsmath}
\usepackage{graphicx}
\usepackage{dcolumn}
\usepackage{bm}
\usepackage{epsfig}
\usepackage[T1]{fontenc}
\usepackage{ae,aecompl}

\begin{document}

\title{Prandtl number effects in MRT Lattice Boltzmann models for shocked and unshocked compressible fluids}
\author{Feng Chen$^1$, Aiguo Xu$^2$\footnote{
Corresponding author. E-mail: Xu\_Aiguo@iapcm.ac.cn}, Guangcai Zhang$^2$,
Yingjun Li$^1$\footnote{%
Corresponding author. E-mail: lyj@aphy.iphy.ac.cn}}
\affiliation{1,State Key Laboratory for GeoMechanics and Deep Underground Engineering, \\
China University of Mining and Technology (Beijing), Beijing 100083 \\
2, National Key Laboratory of Computational Physics, Institute of Applied
Physics and Computational Mathematics, P. O. Box 8009-26, Beijing 100088,
P.R.China}
\date{\today }

\begin{abstract}
For compressible fluids under shock wave reaction, we have proposed two Multiple-Relaxation-Time
(MRT) Lattice Boltzmann (LB) models [F. Chen,  et al, EPL \textbf{90} (2010) 54003;
 Phys. Lett. A \textbf{375} (2011) 2129.].  In this paper,
we construct a new MRT Lattice Boltzmann model which is not only for
the shocked compressible fluids, but also for the unshocked
compressible fluids. To make the model work for unshocked
compressible fluids, a key step is to modify the collision operators
of energy flux so that the viscous coefficient in momentum equation
is consistent with that in energy equation even in the unshocked
system. The unnecessity of the modification for systems under strong
shock is analyzed. The model is validated by some well-known
benchmark tests, including (i) thermal Couette flow, (ii) Riemann
problem, (iii) Richtmyer-Meshkov instability. The first system is
unshocked and the latter two are shocked. In all the three systems,
the Prandtl numbers effects are checked. Satisfying agreements are
obtained between new model results and analytical ones or other
numerical results.
\end{abstract}

 \maketitle

\section{Introduction}

In recent years, the Lattice Boltzmann (LB) method has attracted much
attention as a powerful tool in direct numerical simulation of fluid flows%
\cite{2,BSV,XGL1}. Unlike traditional computational fluid dynamics
methods which solve macroscopic governing equations, the LB method
employs the discrete Boltzmann equation which describes the fluid on
the mesoscale level. This kinetic nature provides the LB method with
essential physics.

However, there are also some limitations that restrict the
applications of traditional LB method, such as the numerical
stability problem, the fixed Prandtl number, and so on. The
stability problem has been partly addressed by a number of
techniques, such as the entropic method\cite{16+,17+},
flux-limiter\cite{Sofonea1} and dissipation\cite{43,Brownlee1}
techniques. Besides these techniques, an
effective method is the Multiple Relaxation Time (MRT) LB method\cite%
{HSB,13,duality}, which employs multiple relaxation parameters in
the collision step, instead of the commonly used Single Relaxation
Time (SRT) collision. The flexibility gained from the MRT collision
can be used to improve the stability property and overcome the fixed
Prandtl number problem.

To the authors' knowledge, most of the existing MRT LB models work
only for isothermal system\cite{17,DuShi,McCracken,guo1}, to cite
but a few. To simulate system with temperature field, Luo, et
al.\cite{19} suggested a hybrid thermal MRT LB model, in which the
mass and momentum equations are solved by the MRT model, whereas the
diffusion-advection equation for the temperature is solved by Finite
Difference (FD) technique or other means. Guo, et al.\cite{guo2}
proposed a coupling MRT LB model for thermal flows with viscous heat
dissipation and compression work. Mezrhab, et al.\cite{Mezrhab}
proposed a double MRT LB method, where MRT-D2Q9 model and the
MRT-D2Q5 model are used to solve the flow and the temperature
fields, respectively.

Besides the models mentioned above, we have proposed two MRT finite
difference Lattice Boltzmann models for compressible fluids under
shock in previous work\cite{chenepl,chenpla}. Numerical experiments
showed that compressible flows with strong shocks can be well
simulated by these models. In this paper, we further propose a new
MRT Lattice Boltzmann model, which is not only for the shocked
compressible fluids, but also for the unshocked compressible fluids.
The rest of the paper is organized as follows: In Sec. II, we
present the MRT LB model. The von Neumann stability analysis is
given in Section III. Simulation results are presented and analyzed
in Section IV. Section V makes the conclusion.


\section{Description of the MRT LB model}

In the MRT LB method, the evolution of the distribution function
$f_{i}$ is governed by the following equation
\begin{equation}
\frac{\partial f_{i}}{\partial t}+v_{i\alpha }\frac{\partial f_{i}}{\partial
x_{\alpha }}=-\mathbf{M}_{il}^{-1}\hat{\mathbf{S}}_{lk}(\hat{f}_{k}-\hat{f}%
_{k}^{eq})\text{,}  \label{3}
\end{equation}%
where $v_{i\alpha}$ is the discrete particle velocity, $i=1$,$\ldots$ ,$N$, $%
N$ is the number of discrete velocities, the subscript $\alpha $
indicates $x$ or $y$. The variable $t$ is time, $x_{\alpha }$ is the
spatial coordinate. The matrix $\hat{\mathbf{S}}=\mathbf{MSM}^{-1}=diag(s_{1},s_{2},%
\cdots ,s_{N})$ is the diagonal relaxation matrix, $f_{i}$ and $\hat{f}_{i}$
are the particle distribution function in the velocity space and the kinetic
moment space respectively, $\hat{f}_{i}=m_{ij}f_{j}$, $m_{ij}$ is an element
of the transformation matrix $\mathbf{M}$. Obviously, the mapping between
moment space and velocity space is defined by the linear transformation $%
\mathbf{M}$, i.e., $\hat{\mathbf{f}}=\mathbf{Mf}$, $\mathbf{f=M}^{-1}\hat{%
\mathbf{f}}$, where the bold-face symbols denote N-dimensional column
vectors, e.g., $\mathbf{f}=( f_{1},f_{2},\cdots ,f_{N})^{T}$, $\hat{\mathbf{f%
}}=( \hat{f}_{1},\hat{f}_{2},\cdots ,\hat{f}_{N})^{T}$, $\mathbf{M}=(
m_{1},m_{2},\cdots ,m_{N}) ^{T} $, $m_{i}=(m_{i1},m_{i2},\cdots ,m_{iN})$. $%
\hat{f}_{i}^{eq}$\ is the equilibrium value of the moment $\hat{f}_{i}$.

\begin{figure}[tbp]
\center\includegraphics*[width=0.40\textwidth]{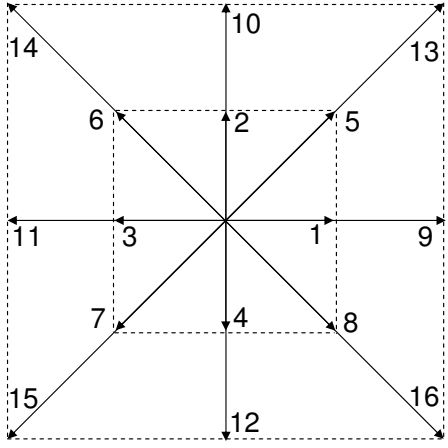} \caption{
Schematics of $\mathbf{v}_{i}$ for the discrete velocity model.}
\end{figure}
We construct a two-dimensional MRT LB model based on a $16$%
-discrete-velocity model (see Fig. 1):
\begin{equation*}
\left( v_{i1,}v_{i2}\right) =\left\{
\begin{array}{cc}
\mathbf{cyc}:\left( \pm 1,0\right) , & \text{for }1\leq i\leq 4, \\
\left( \pm 1,\pm 1\right) , & \text{for }5\leq i\leq 8, \\
\mathbf{cyc}:\left( \pm 2,0\right) , & \text{for }9\leq i\leq 12, \\
\left( \pm 2,\pm 2\right) , & \text{for }13\leq i\leq 16,%
\end{array}%
\right.
\end{equation*}%
where \textbf{cyc} indicates the cyclic permutation.

The transformation matrix $\mathbf{M}$ is constructed according to the
irreducible representation bases of SO(2) group, and it can be expressed as
follows:
\begin{equation*}
\mathbf{M}=(m_{1},m_{2},\cdots ,m_{16})^{T}\text{,}
\end{equation*}%
where%
\begin{equation*}
m_{1}=(1,1,1,1,1,1,1,1,1,1,1,1,1,1,1,1)\text{,}
\end{equation*}%
\begin{equation*}
m_{2}=(1,0,-1,0,1,-1,-1,1,2,0,-2,0,2,-2,-2,2)\text{,}
\end{equation*}%
\begin{equation*}
m_{3}=(0,1,0,-1,1,1,-1,-1,0,2,0,-2,2,2,-2,-2)\text{,}
\end{equation*}%
\begin{equation*}
m_{4}=(\frac{1}{2},\frac{1}{2},\frac{1}{2},\frac{1}{2}%
,1,1,1,1,2,2,2,2,4,4,4,4)\text{,}
\end{equation*}%
\begin{equation*}
m_{5}=(1,-1,1,-1,0,0,0,0,4,-4,4,-4,0,0,0,0)\text{,}
\end{equation*}%
\begin{equation*}
m_{6}=(0,0,0,0,1,-1,1,-1,0,0,0,0,4,-4,4,-4)\text{,}
\end{equation*}%
\begin{equation*}
m_{7}=(\frac{1}{2},0,-\frac{1}{2},0,1,-1,-1,1,4,0,-4,0,8,-8,-8,8)\text{,}
\end{equation*}%
\begin{equation*}
m_{8}=(0,\frac{1}{2},0,-\frac{1}{2},1,1,-1,-1,0,4,0,-4,8,8,-8,-8)\text{,}
\end{equation*}%
\begin{equation*}
m_{9}=(1,0,-1,0,-2,2,2,-2,8,0,-8,0,-16,16,16,-16)\text{,}
\end{equation*}%
\begin{equation*}
m_{10}=(0,-1,0,1,2,2,-2,-2,0,-8,0,8,16,16,-16,-16)\text{,}
\end{equation*}%
\begin{equation*}
m_{11}=(\frac{1}{4},\frac{1}{4},\frac{1}{4},\frac{1}{4}%
,1,1,1,1,4,4,4,4,16,16,16,16)\text{,}
\end{equation*}%
\begin{equation*}
m_{12}=(1,1,1,1,-4,-4,-4,-4,16,16,16,16,-64,-64,-64,-64)\text{,}
\end{equation*}%
\begin{equation*}
m_{13}=(1,-1,1,-1,0,0,0,0,16,-16,16,-16,0,0,0,0)\text{,}
\end{equation*}%
\begin{equation*}
m_{14}=(0,0,0,0,2,-2,2,-2,0,0,0,0,32,-32,32,-32)\text{,}
\end{equation*}%
\begin{equation*}
m_{15}=(1,0,-1,0,-4,4,4,-4,32,0,-32,0,-128,128,128,-128)\text{,}
\end{equation*}%
\begin{equation*}
m_{16}=(0,-1,0,1,4,4,-4,-4,0,-32,0,32,128,128,-128,-128)\text{.}
\end{equation*}

For two-dimensional compressible models, we have four conserved
moments, density $\rho$, momentums $j_{x}$, $j_{y}$, and energy $e$.
They are denoted by $\hat{f}_{1}$, $\hat{f}_{2}$, $\hat{f}_{3}$ and
$\hat{f}_{4}$,
respectively. Specifically, $\hat{f}_{1}=\rho $, $\hat{f}_{2}=j_{x}$, $\hat{f%
}_{3}=j_{y}$, $\hat{f}_{4}=e=\rho (T+u^{2}/2)$. Using the
Chapman-Enskog expansion\cite{DuShi,McCracken,22} on the two sides
of LB equation, the Navier-Stokes (NS) equations for compressible
fluids can be derived. The equilibria of the nonconserved moments
can be chosen as
\begin{subequations}
\begin{equation}
\hat{f}_{5}^{eq}=(j_{x}^{2}-j_{y}^{2})/\rho ,  \label{14a}
\end{equation}%
\begin{equation}
\hat{f}_{6}^{eq}=j_{x}j_{y}/\rho ,  \label{14b}
\end{equation}%
\begin{equation}
\hat{f}_{7}^{eq}=(e+\rho RT)j_{x}/\rho ,  \label{14c}
\end{equation}%
\begin{equation}
\hat{f}_{8}^{eq}=(e+\rho RT)j_{y}/\rho ,  \label{14d}
\end{equation}%
\begin{equation}
\hat{f}_{9}^{eq}=(j_{x}^{2}-3j_{y}^{2})j_{x}/\rho ^{2},  \label{14e}
\end{equation}%
\begin{equation}
\hat{f}_{10}^{eq}=(3j_{x}^{2}-j_{y}^{2})j_{y}/\rho ^{2},  \label{14f}
\end{equation}%
\begin{equation}
\hat{f}_{11}^{eq}=2e^{2}/\rho -(j_{x}^{2}+j_{y}^{2})^{2}/4\rho ^{3},
\label{14g}
\end{equation}%
\begin{equation}
\hat{f}_{13}^{eq}=(6\rho e-2j_{x}^{2}-2j_{y}^{2})(j_{x}^{2}-j_{y}^{2})/\rho
^{3},  \label{14h}
\end{equation}%
\begin{equation}
\hat{f}_{14}^{eq}=(6\rho e-2j_{x}^{2}-2j_{y}^{2})j_{x}j_{y}/\rho ^{3}\text{.}
\label{14i}
\end{equation}%
\end{subequations}
The recovered NS equations are as follows:
\begin{subequations}\label{Eq3}
\begin{equation}
\frac{\partial \rho }{\partial t}+\frac{\partial j_{x}}{\partial x}+\frac{%
\partial j_{y}}{\partial y}=0,  \label{15a}
\end{equation}%
\begin{equation}
\frac{\partial j_{x}}{\partial t}+\frac{\partial }{\partial x}\left(
j_{x}^{2}/\rho \right) +\frac{\partial }{\partial y}\left( j_{x}j_{y}/\rho
\right) =-\frac{\partial P}{\partial x}+\frac{\partial }{\partial x}[\mu
_{s}(\frac{\partial u_{x}}{\partial x}-\frac{\partial u_{y}}{\partial y})]+%
\frac{\partial }{\partial y}[\mu _{v}(\frac{\partial u_{y}}{\partial x}+%
\frac{\partial u_{x}}{\partial y})],  \label{15b}
\end{equation}%
\begin{equation}
\frac{\partial j_{y}}{\partial t}+\frac{\partial }{\partial x}\left(
j_{x}j_{y}/\rho \right) +\frac{\partial }{\partial y}\left( j_{y}^{2}/\rho
\right) =-\frac{\partial P}{\partial y}+\frac{\partial }{\partial x}[\mu
_{v}(\frac{\partial u_{y}}{\partial x}+\frac{\partial u_{x}}{\partial y})]-%
\frac{\partial }{\partial y}[\mu _{s}(\frac{\partial u_{x}}{\partial x}-%
\frac{\partial u_{y}}{\partial y})],  \label{15c}
\end{equation}%
\begin{eqnarray}
&&\frac{\partial e}{\partial t}+\frac{\partial }{\partial x}[(e+P)j_{x}/\rho
]+\frac{\partial }{\partial y}[(e+P)j_{y}/\rho ]  \notag \\
&=&\frac{\partial }{\partial x}[\lambda _{1}(R\frac{\partial T}{\partial x}+%
\frac{1}{2}(u_{y}\frac{\partial u_{y}}{\partial x}+u_{x}\frac{\partial u_{x}%
}{\partial x}-u_{x}\frac{\partial u_{y}}{\partial y}+u_{y}\frac{\partial
u_{x}}{\partial y}))]  \notag \\
&&+\frac{\partial }{\partial y}[\lambda _{2}(R\frac{\partial T}{\partial y}+%
\frac{1}{2}(u_{x}\frac{\partial u_{x}}{\partial y}-u_{y}\frac{\partial u_{x}%
}{\partial x}+u_{x}\frac{\partial u_{y}}{\partial x}+u_{y}\frac{\partial
u_{y}}{\partial y}))],  \label{15d}
\end{eqnarray}%
where $\mu _{s}=$ $\rho RT/s_{5}$, $\mu _{v}=$ $\rho RT/s_{6}$, $\lambda
_{1}=2\rho RT/s_{7}$, $\lambda _{2}=2\rho RT/s_{8}$.

When $\mu _{s}=$\ $\mu _{v}=\mu $,\ $\lambda _{1}=\lambda _{2}=\lambda $,
the above NS equations reduce to
\end{subequations}
\begin{subequations}
\begin{equation}
\frac{\partial \rho }{\partial t}+\frac{\partial j_{\alpha }}{\partial
x_{\alpha }}=0,  \label{16a}
\end{equation}%
\begin{equation}
\frac{\partial j_{\alpha }}{\partial t}+\frac{\partial \left( j_{\alpha
}j_{\beta }/\rho \right) }{\partial x_{\beta }}=-\frac{\partial P}{\partial
x_{\alpha }}+\frac{\partial }{\partial x_{\beta }}[\mu (\frac{\partial
u_{\alpha }}{\partial x_{\beta }}+\frac{\partial u_{\beta }}{\partial
x_{\alpha }}-\frac{\partial u_{\chi }}{\partial x_{\chi }}\delta _{\alpha
\beta })],  \label{16b}
\end{equation}%
\begin{equation}
\frac{\partial e}{\partial t}+\frac{\partial }{\partial x_{\alpha }}%
[(e+P)j_{\alpha }/\rho ]=\frac{\partial }{\partial x_{\alpha }}[\lambda (R%
\frac{\partial T}{\partial x_{\alpha }}+\frac{1}{2}u_{\beta }(\frac{\partial
u_{\alpha }}{\partial x_{\beta }}+\frac{\partial u_{\beta }}{\partial
x_{\alpha }}-\frac{\partial u_{\chi }}{\partial x_{\chi }}\delta _{\alpha
\beta }))]\text{.}  \label{16c}
\end{equation}

It should be pointed out that, the viscous coefficient in the energy
equation \eqref{16c} is not consistent with that in the momentum
equation \eqref{16b}. Motivated by the idea of Guo et al.
\cite{guo2}, the collision operators of the moments related to the
energy flux are modified:
\end{subequations}
\begin{equation*}
\hat{\mathbf{S}}_{77}(\hat{f}_{7}-\hat{f}_{7}^{eq})\Rightarrow \hat{\mathbf{S%
}}_{77}(\hat{f}_{7}-\hat{f}_{7}^{eq})+(s_{7}/s_{5}-1)\rho Tu_{x}(\frac{%
\partial u_{x}}{\partial x}-\frac{\partial u_{y}}{\partial y}%
)+(s_{7}/s_{6}-1)\rho Tu_{y}(\frac{\partial u_{y}}{\partial x}+\frac{%
\partial u_{x}}{\partial y})\text{,}
\end{equation*}%
\begin{equation*}
\hat{\mathbf{S}}_{88}(\hat{f}_{8}-\hat{f}_{8}^{eq})\Rightarrow \hat{\mathbf{S%
}}_{88}(\hat{f}_{8}-\hat{f}_{8}^{eq})+(s_{8}/s_{6}-1)\rho Tu_{x}(\frac{%
\partial u_{y}}{\partial x}+\frac{\partial u_{x}}{\partial y}%
)+(s_{8}/s_{5}-1)\rho Tu_{y}(\frac{\partial u_{x}}{\partial x}-\frac{%
\partial u_{y}}{\partial y})\text{.}
\end{equation*}

With this modification, we are able to get the following
thermohydrodynamic equations:
\begin{subequations}\label{Eq5}
\begin{equation}
\frac{\partial \rho }{\partial t}+\frac{\partial j_{\alpha }}{\partial
x_{\alpha }}=0,
\end{equation}%
\begin{equation}
\frac{\partial j_{\alpha }}{\partial t}+\frac{\partial \left( j_{\alpha
}j_{\beta }/\rho \right) }{\partial x_{\beta }}=-\frac{\partial P}{\partial
x_{\alpha }}+\frac{\partial }{\partial x_{\beta }}[\mu (\frac{\partial
u_{\alpha }}{\partial x_{\beta }}+\frac{\partial u_{\beta }}{\partial
x_{\alpha }}-\frac{\partial u_{\chi }}{\partial x_{\chi }}\delta _{\alpha
\beta })],
\end{equation}%
\begin{equation}
\frac{\partial e}{\partial t}+\frac{\partial }{\partial x_{\alpha }}%
[(e+P)j_{\alpha }/\rho ]=\frac{\partial }{\partial x_{\alpha
}}[\lambda R\frac{\partial T}{\partial x_{\alpha }}+\mu u_{\beta
}(\frac{\partial u_{\alpha }}{\partial x_{\beta }}+\frac{\partial
u_{\beta }}{\partial x_{\alpha }}-\frac{\partial u_{\chi }}{\partial
x_{\chi }}\delta _{\alpha \beta })]\text{.}
\end{equation}
\end{subequations}

This modification method is also suitable for our previous MRT
models\cite{chenepl,chenpla}. The definitions of
$\hat{f}_{12}^{eq}$,\ $\hat{f}_{15}^{eq}$,\ $\hat{f}_{16}^{eq} $
have no effect on macroscopic equations, so the choices of the three
moments are flexible. Now we give three different and typical formations: $%
\hat{f}_{12}^{eq}=\hat{f}_{15}^{eq}=\hat{f}_{16}^{eq}=0$ (version 1); $\hat{f%
}_{12}^{eq}=0$,$\ \hat{f}_{15}^{eq}=\rho u_{x}(-4+10T+5u_{x}^{2}-5u_{y}^{2})$%
, $\hat{f}_{16}^{eq}=\rho u_{y}(4-10T+5u_{x}^{2}-5u_{y}^{2})$ (version 2); $%
\hat{f}_{12}^{eq}=M_{12i}f_{i}^{\max }$, $\hat{f}_{15}^{eq}=M_{15i}f_{i}^{%
\max }$, $\hat{f}_{16}^{eq}=M_{16i}f_{i}^{\max }$, $f_{i}^{\max }=\rho
/(2\pi T)\exp (-(v_{i\alpha }-u_{\alpha })^{2}/(2T))$ (version 3). In the
second version, the MRT model reduces to the usual lattice BGK model in ref.%
\cite{Chen1994} which uses a higher-order velocity expansion for
Maxwellian-type equilibrium distribution, if all the relaxation
parameters
are set to be a single relaxation frequency $s$, namely $\mathbf{S}=s\mathbf{%
I}$.

\section{Stability Analysis}

In this section, the von Neumann stability analysis\cite{chenpla} on
the new MRT LB model is performed. In the stability analysis, we
write the solution of FD LB equation in the form of Fourier series.
If all the eigenvalues of the coefficient matrix are less than 1,
the algorithm is stable. Coefficient matrix $G_{ij}$ of the
unmodified model can be expressed as follows,
\begin{align}
G_{ij}& =\delta _{ij}-\frac{v_{i\alpha }\Delta t}{2\Delta x_{\alpha }}(e^{%
\mathbf{i}k_{\alpha }\Delta x_{\alpha }}-e^{-\mathbf{i}k_{\alpha
}\Delta x_{\alpha }})\delta _{ij}+\frac{1}{2}(\frac{v_{i\alpha
}\Delta t}{\Delta
x_{\alpha }})^{2}(e^{\mathbf{i}k_{\alpha }\Delta x_{\alpha }}-2  \notag \\
& +e^{-\mathbf{i}k_{\alpha }\Delta x_{\alpha }})\delta _{ij}-\Delta t\mathbf{%
M}_{il}^{-1}\hat{\mathbf{S}}_{lk}(\frac{\partial \hat{f}_{k}}{\partial f_{j}}%
-\frac{\partial \hat{f}_{k}^{eq}}{\partial f_{j}})\text{,}
\end{align}%
where
\begin{equation*}
\hat{f}_{k}=\mathbf{M}_{kp}f_{p}\text{,}
\end{equation*}%
\begin{equation}
\frac{\partial \hat{f}_{k}^{eq}}{\partial f_{j}}=\frac{\partial \hat{f}%
_{k}^{eq}}{\partial \rho }\frac{\partial \rho }{\partial f_{j}}+\frac{%
\partial \hat{f}_{k}^{eq}}{\partial T}\frac{\partial T}{\partial f_{j}}+%
\frac{\partial \hat{f}_{k}^{eq}}{\partial u_{\alpha }}\frac{\partial
u_{\alpha }}{\partial f_{j}}\text{.}
\end{equation}%
Because of the modification of the collision operators, the
coefficient matrix $G_{ij}$ corresponding to energy flux should be
replaced by
\begin{align}
G_{ij}& =\delta _{ij}-\frac{v_{i\alpha }\Delta t}{2\Delta x_{\alpha }}(e^{%
\mathbf{i}k_{\alpha }\Delta x_{\alpha }}-e^{-\mathbf{i}k_{\alpha
}\Delta x_{\alpha }})\delta _{ij}+\frac{1}{2}(\frac{v_{i\alpha
}\Delta t}{\Delta
x_{\alpha }})^{2}(e^{\mathbf{i}k_{\alpha }\Delta x_{\alpha }}-2  \notag \\
& +e^{-\mathbf{i}k_{\alpha }\Delta x_{\alpha }})\delta _{ij}-\Delta t\mathbf{%
M}_{i7}^{-1}\hat{\{\mathbf{S}}_{77}(\frac{\partial
\hat{f}_{7}}{\partial
f_{j}}-\frac{\partial \hat{f}_{7}^{eq}}{\partial f_{j}})+\frac{\partial }{%
\partial f_{j}}[(s_{7}/s_{5}-1)\rho Tu_{x}(\frac{\partial u_{x}}{\partial x}-%
\frac{\partial u_{y}}{\partial y})]  \notag \\
& +\frac{\partial }{\partial f_{j}}[(s_{7}/s_{6}-1)\rho Tu_{y}(\frac{%
\partial u_{y}}{\partial x}+\frac{\partial u_{x}}{\partial y})]\}\text{,}
\end{align}%
and%
\begin{align}
G_{ij}& =\delta _{ij}-\frac{v_{i\alpha }\Delta t}{2\Delta x_{\alpha }}(e^{%
\mathbf{i}k_{\alpha }\Delta x_{\alpha }}-e^{-\mathbf{i}k_{\alpha
}\Delta x_{\alpha }})\delta _{ij}+\frac{1}{2}(\frac{v_{i\alpha
}\Delta t}{\Delta
x_{\alpha }})^{2}(e^{\mathbf{i}k_{\alpha }\Delta x_{\alpha }}-2  \notag \\
& +e^{-\mathbf{i}k_{\alpha }\Delta x_{\alpha }})\delta _{ij}-\Delta t\mathbf{%
M}_{i8}^{-1}\hat{\{\mathbf{S}}_{88}(\frac{\partial
\hat{f}_{8}}{\partial
f_{j}}-\frac{\partial \hat{f}_{8}^{eq}}{\partial f_{j}})+\frac{\partial }{%
\partial f_{j}}[(s_{8}/s_{6}-1)\rho Tu_{x}(\frac{\partial u_{y}}{\partial x}+%
\frac{\partial u_{x}}{\partial y})]  \notag \\
& +\frac{\partial }{\partial f_{j}}[(s_{8}/s_{5}-1)\rho Tu_{y}(\frac{%
\partial u_{x}}{\partial x}-\frac{\partial u_{y}}{\partial y})]\}\text{.}
\end{align}

We conduct a quantitative analysis using the software, Mathematica.
In Fig.2 we show some stability comparisons for the new MRT model,
its SRT counterpart and our previous model in
refs.\cite{chenepl,chenpla}. The abscissa is for $kdx$, and the
vertical axis is for $|\omega |_{max}$ which
is the largest eigenvalue of coefficient matrix $G_{ij}$. Grid sizes are $%
dx=dy=10^{-3}$, and time step is $dt=10^{-5}$, the relaxation
frequency in SRT is $s=10^{5}$. The other parameters in stability
analysis are chosen as follows: (a), $(\rho ,u_{1},u_{2},T)$ =
$(2.0,2.0,0.0,2.0)$, the collision
parameters in MRT are $s_{i}=10^{5},i=1,\cdots ,16$, the Mach number is 1 ($%
Ma=u/\sqrt{2T}=2/2$); (b), $(\rho ,u_{1},u_{2},T)$ =
$(2.0,6.0,0.0,2.0)$, the collision parameters in the three versions
are $s_{9}=10^{3}$, those in
model\cite{chenepl} are $s_{10}=5\times 10^{4}$, $s_{11}=2\times 10^{4}$, $%
s_{13}=1.5\times 10^{4}$, and those in \cite{chenpla} are
$s_{9}=8\times
10^{3}$, $s_{13}=7\times 10^{4}$, $s_{14}=5\times 10^{4}$, the others are $%
10^{5}$, and the Mach number is 3.0; (c), $(\rho ,u_{1},u_{2},T)$ = $%
(2.0,10.0,0.0,2.0)$, the collision parameters in the three versions are $%
s_{9}=1.2\times 10^{4}$, $s_{13}=10^{2}$, $s_{14}=5\times 10^{4}$, $%
s_{15}=1.5\times 10^{3}$, those in model\cite{chenepl} are
$s_{13}=1.5\times
10^{4}$, and those in model\cite{chenpla} are $s_{9}=2\times 10^{3}$, $%
s_{13}=6.1\times 10^{4}$, $s_{14}=s_{15}=3\times 10^{4}$, the others are $%
10^{5}$, and the Mach number is 5; (d), $(\rho ,u_{1},u_{2},T)$ = $%
(2.0,12.0,0.0,2.0)$, the collision parameters in the three versions are $%
s_{9}=10^{4}$, $s_{13}=10^{2}$, $s_{14}=6\times 10^{4}$,
$s_{15}=1.5\times
10^{3}$, and those in model\cite{chenepl} are $s_{11}=2\times 10^{4}$, $%
s_{13}=1.5\times 10^{4}$, and those in \cite{chenpla} are $s_{9}=10^{3}$, $%
s_{13}=5\times 10^{3}$, $s_{14}=s_{15}=3\times 10^{4}$, the others are $%
10^{5}$, and the Mach number is 6. In case (a), the MRT and SRT have
the same stability; with the increase of Mach number (case (b) and
(c)), the MRT models are stable, while the SRT version is not; if
further increases the Mach number, MRT models also encounter
instability problem (case (d)). It is clear that, by choosing
appropriate collision parameters, the stability of MRT can be much
better than the SRT. Three versions of the new MRT model do not show
large differences in numerical stability.
\begin{figure}[tbp]
\center\includegraphics*%
[bbllx=15pt,bblly=18pt,bburx=335pt,bbury=230pt,width=0.8\textwidth]{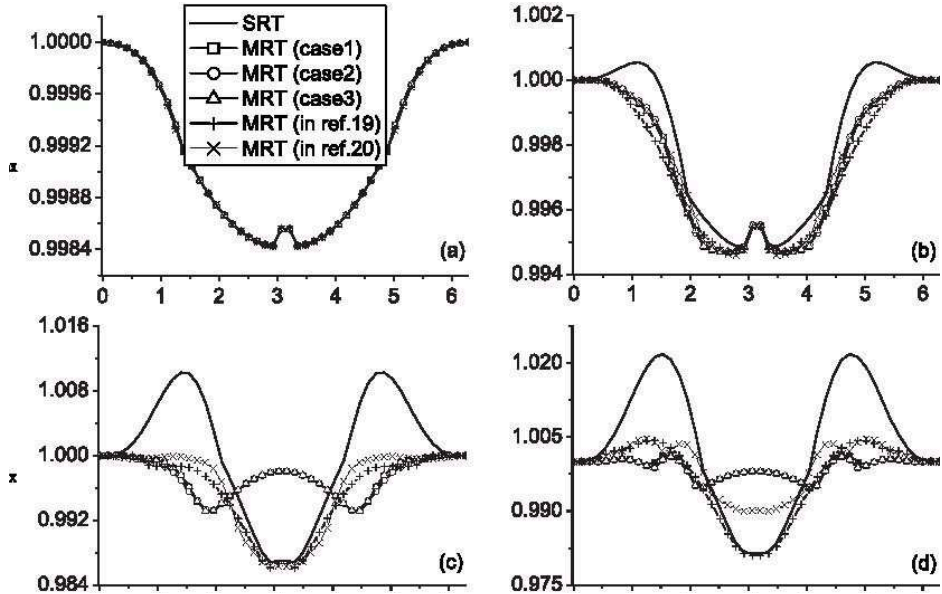}
\caption{Stability comparison for the new MRT model and its SRT
counterpart.}
\end{figure}

\section{Numerical Simulations}

In this section, we study the following problems using the modified
MRT LB model: Couette flow, One-dimensional Riemann problem, and
Richtmyer-Meshkov instability. We work in a frame where the constant
$R=1$.

\subsection{Unshocked compressible fluids}

Here we conduct a series of numerical simulations of Couette flow.
The aims of simulation of Couette flow are twofold. At first, we
prove the Maxwellian property of the discrete equilibrium functions.
Consider a viscous fluid flow between two parallel flat plates,
moving in the opposite directions, $U_{wr}=-U_{wl}=0.2$, where
subscripts $wr$ and $wl$ indicate the walls in the right and left
sides. The initial state of the fluid is $\rho =1$, $T=1$, $U=0$.
The temperatures of walls are $T_{wr}=T_{wl}=1$. Near the walls, we
adopt the
diffuse reflection boundary conditions proposed by Sofonea, et al\cite%
{diffuse}. In the other two boundaries the periodic boundary condition is
adopted. In the diffuse reflection boundary, the particles leaving the wall
are assumed to follow the Maxwellian distribution. Following the
discretization of the velocity space, in the FD LB model the Maxwellian
distribution function is replaced by the equilibrium distribution function.
\begin{figure}[tbp]
\begin{center}
\includegraphics[bbllx=21pt,bblly=25pt,bburx=450pt,bbury=240pt,
scale=0.9]{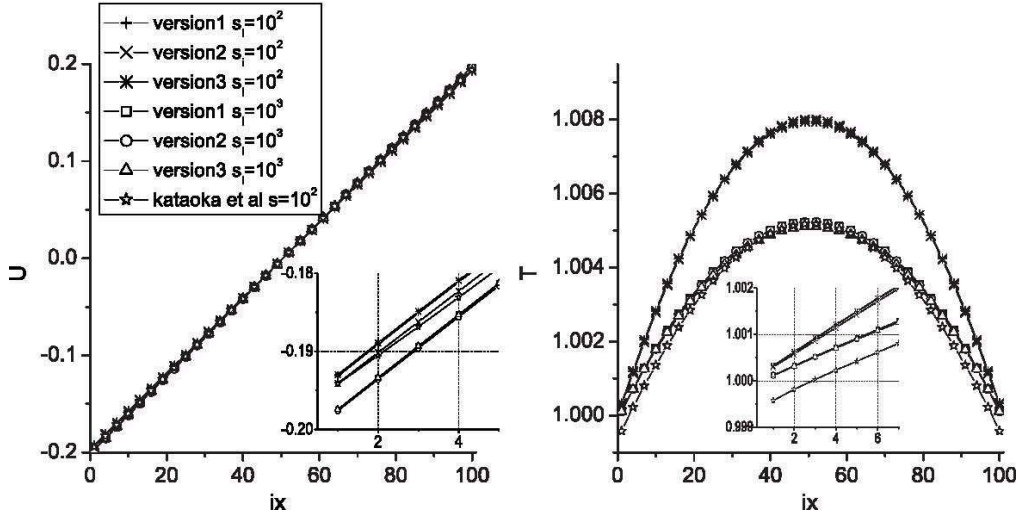}
\end{center}
\caption{Slip velocity and temperature jump simulated with the three
versions and the model proposed by Kataoka, et al.}
\end{figure}

Fig. 3 shows the velocity and temperature profiles simulated with
the three versions of this proposed model and the model proposed by
Kataoka, et al\cite{Kataoka}. The abscissa $ix$ is the index of
lattice node in the $x$- directions, and the vertical axes are
velocity $u$ and temperature $T$, respectively. The parameters are
$dx=dy=0.01$, $dt=10^{-4}$, $NX\times NY=100\times 5$. The diffuse
reflection boundary conditions work well with our model, the slip
velocity and temperature jump near the walls are clearly seen, and
increase with Knudsen number. While it fails to work for the model
by Kataoka et al, because the temperature near the wall is lower
than the wall temperature, which is contrary to physical idea. In
Couette flow the fluid at the walls should have a higher temperature
than the walls themselves, because of the heat generated by the
viscous flow. We think this contradiction is caused from the
equilibrium distribution function in their model which is not a
Taylor expansion of the Maxwellian. So it departs from the basic
assumption of diffuse reflection boundary. None of the three
versions violates the basic assumption and destroys the Maxwellian
property of the discrete equilibrium functions.

\begin{figure}[tbp]
\begin{center}
\includegraphics[bbllx=18pt,bblly=18pt,bburx=370pt,bbury=210pt,
scale=0.9]{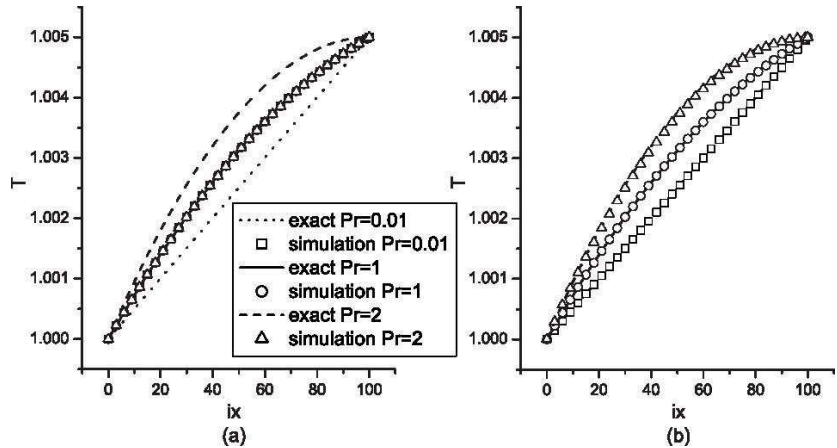}
\end{center}
\caption{Effects of heat conductivity on temperature profiles of
Couette flow. (a) corresponds to the unmodified model(version 1),
and (b) corresponds
to the modified model. $Pr=0.01$, $Pr=1$ and $Pr=2$ correspond to $s_{7}=s_{8}=10$, $s_{7}=s_{8}=10^3$, and $%
s_{7}=s_{8}=2 \times 10^3$, respectively (other collision parameters
are $10^3$).}
\end{figure}
\begin{figure}[tbp]
\begin{center}
\includegraphics[bbllx=18pt,bblly=18pt,bburx=380pt,bbury=210pt,
scale=0.9]{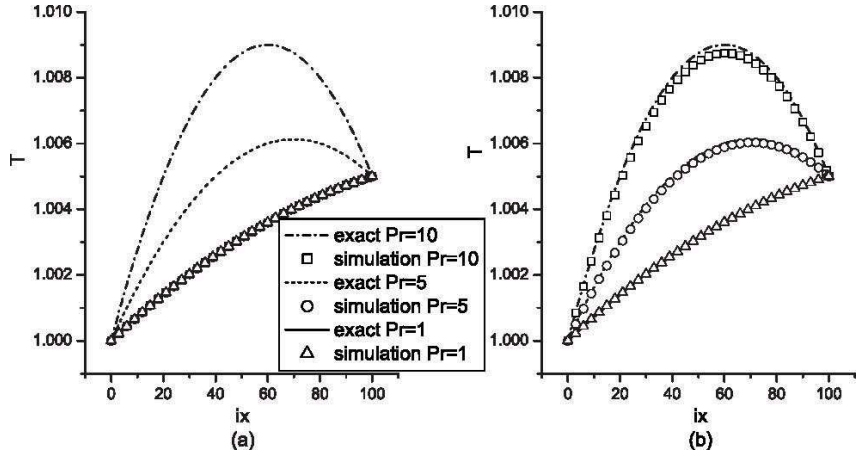}
\end{center}
\caption{Effects of viscosity on temperature profiles of Couette
flow. (a) corresponds to the unmodified model(version 1), and (b)
corresponds
to the modified model. $Pr=10$, $Pr=5$ and $Pr=1$ correspond to $s_{5}=s_{6}=10^2$, $s_{5}=s_{6}=2 \times 10^2$, and $%
s_{5}=s_{6}=10^3$, respectively (other collision parameters are
$10^3$).}
\end{figure}

Secondly, we will compare the ability of the unmodified model and
the modified model for the unshocked compressible fluids. In the
simulation, the left wall is fixed and the right wall moves at speed
$U=0.1$. The simulation results are compared with the analytical
solution:
\begin{equation*}
T=T_{1}+(T_{2}-T_{1})\frac{x}{H}+\frac{\mu }{2\lambda }U^{2}\frac{x}{H}(1-%
\frac{x}{H})\text{,}
\end{equation*}
where $T_{1}$ and $T_{2}$ are the left and right wall's temperatures
($T_{1}=1$, $T_{2}=1.005$), $H$ is the width of the channel. Other
parameters remain unchanged. Periodic boundary conditions are
applied to the bottom and top boundaries, and the left and right
walls adopt the nonequilibrium extrapolation method. Fig. 4 and Fig.
5 show the temperature profiles of Couette flow simulated with the
unmodified model (version 1) and its modified version. In Fig. 4, we
fix viscosity coefficient $s_{5}=s_{6}=10^3$, and change the thermal
conductivity $s_{7}=s_{8}$ from $10$ to $2 \times 10^3$. On the
contrary, we fix thermal conductivity $s_{7}=s_{8}=10^3$, and change
the viscosity $s_{5}=s_{6}$ from $10^2$ to $10^3$ in Fig. 5. (a)
corresponds to the unmodified model (version 1), and (b) corresponds
to the modified model. It is clearly shown that the simulation
results of modified model are in agreement with the analytical
solutions, and the Prandtl number effects on unshocked compressible
fluids are successfully captured by the modified model, but not by
the unmodified model.

\subsection{Shocked compressible fluids}

(a) Riemann problem

Here we construct a high Mach number shock tube problem with the
initial condition,
\begin{equation}
\left\{
\begin{array}{cc}
(\rho ,u_{1},u_{2},T)|_{L}=(5.0,45.0,0.0,10.0), & x\leq 0\text{.} \\
(\rho ,u_{1},u_{2},T)|_{R}=(6.0,-20.0,0.0,5.0), & x>0\text{.}%
\end{array}%
\right.
\end{equation}%
The Mach number of the left side is $10.1$
($Ma=u/\sqrt{2T}=45/\sqrt{20}$), and the right is $6.3$
($Ma=u/\sqrt{2T}=20/\sqrt{10}$). Figure 6 shows the comparison of LB
results and exact solutions at $t=0.018$, where the parameters are $dx=dy=0.003$, $dt=10^{-5}$, $%
s_{5}=s_{6}=1.5\times 10^{4}$, other values of $s$\ are $10^{5}$.
Squares correspond to simulation results with the unmodified model
(version 1), the circle symbols correspond to the modified MRT
simulation results, and solid lines represent the exact solutions,
respectively. It can be seen that the simulations of the two MRT
models do not show large differences. For shocked compressible
flows, there exist a fast procedure and a slow one. The shock dynamic procedure is fast, while that of heat conduction is slow. In such a case, from the view of macroscopic description, the terms related to viscosity and heat conductivity may be neglected. So, terms related to viscosity and heat conductivity in Eqs.(\ref{Eq3}) and (\ref{Eq5}) are all small terms and make negligible effects. That is the reason why the unmodified model works also well in such cases.
\begin{figure}[tbp]
\begin{center}
\includegraphics[bbllx=18pt,bblly=17pt,bburx=320pt,bbury=230pt,scale=1.1]{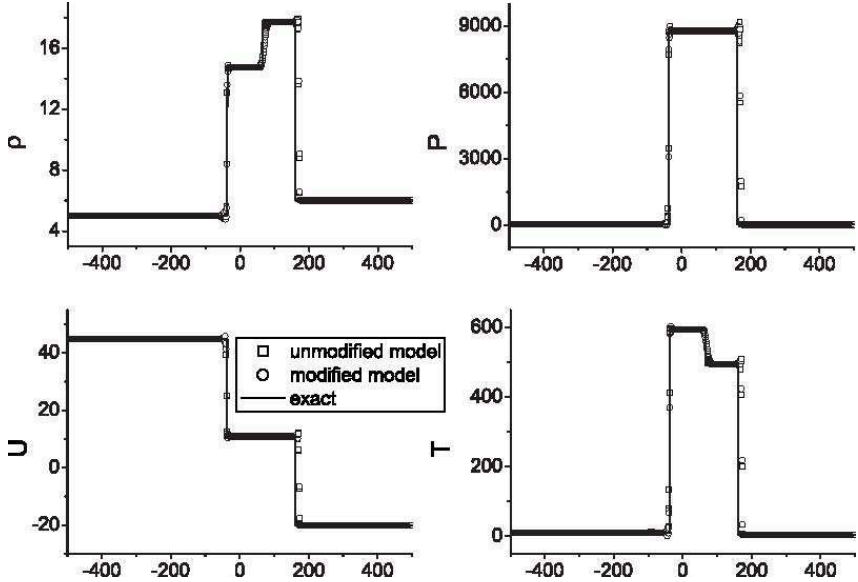}
\end{center}
\caption{LB results and exact solutions for shock tube problem at
time $t=0.018$. $\protect\rho$: density, $P$: pressure, $U$: the
$x-$ component of velocity, $T$: temperature.}
\end{figure}

(b) Richtmyer-Meshkov instability

The Richtmyer-Meshkov (RM) instability\cite{Richtmyer,Meshkov} is a
fundamental fluid instability that develops when an incident shock
wave collides with an interface between two fluids with different
densities. This instability is involved in numerous physical
processes, such as inertial confined fusion, supersonic and
hypersonic combustion, supernova explosion, and so on. RM
instability has attracted considerable attention for several decades
because of its important theoretical and practical significance. To
the best of our knowledge, the research of RM instability by LB
method is still very limited. In this paper, we study the thermal
conductivity and viscosity effects on RM instability with the MRT LB
method.
\begin{figure}[tbp]
\begin{center}
\includegraphics[bbllx=25pt,bblly=60pt,bburx=595pt,bbury=485pt,scale=0.8]{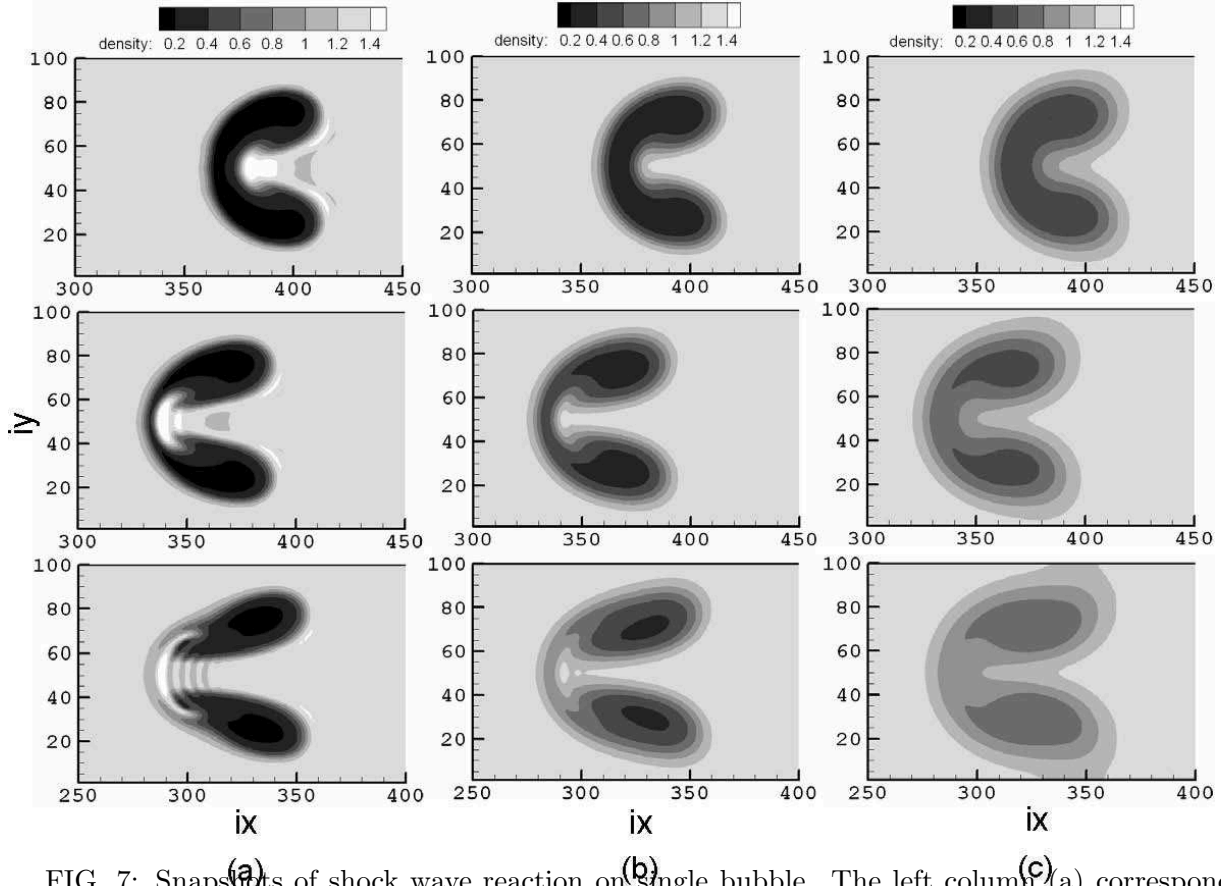}
\end{center}
\caption{Snapshots of shock wave reaction on single bubble. The left
column (a) corresponds to $s_{5}=s_{6}=10^{4}, s_{7}=s_{8}=10^{5}$,
the middle column (b) corresponds to $s_{5}=s_{6}=10^{4},
s_{7}=s_{8}=10^{4}$, and the right column (c) corresponds to
$s_{5}=s_{6}=10^{4}, s_{7}=s_{8}=10^{3}$. From black to white the
grey level corresponds to the increase of density.}
\end{figure}
\begin{figure}[tbp]
\begin{center}
\includegraphics[bbllx=25pt,bblly=60pt,bburx=595pt,bbury=485pt,scale=0.8]{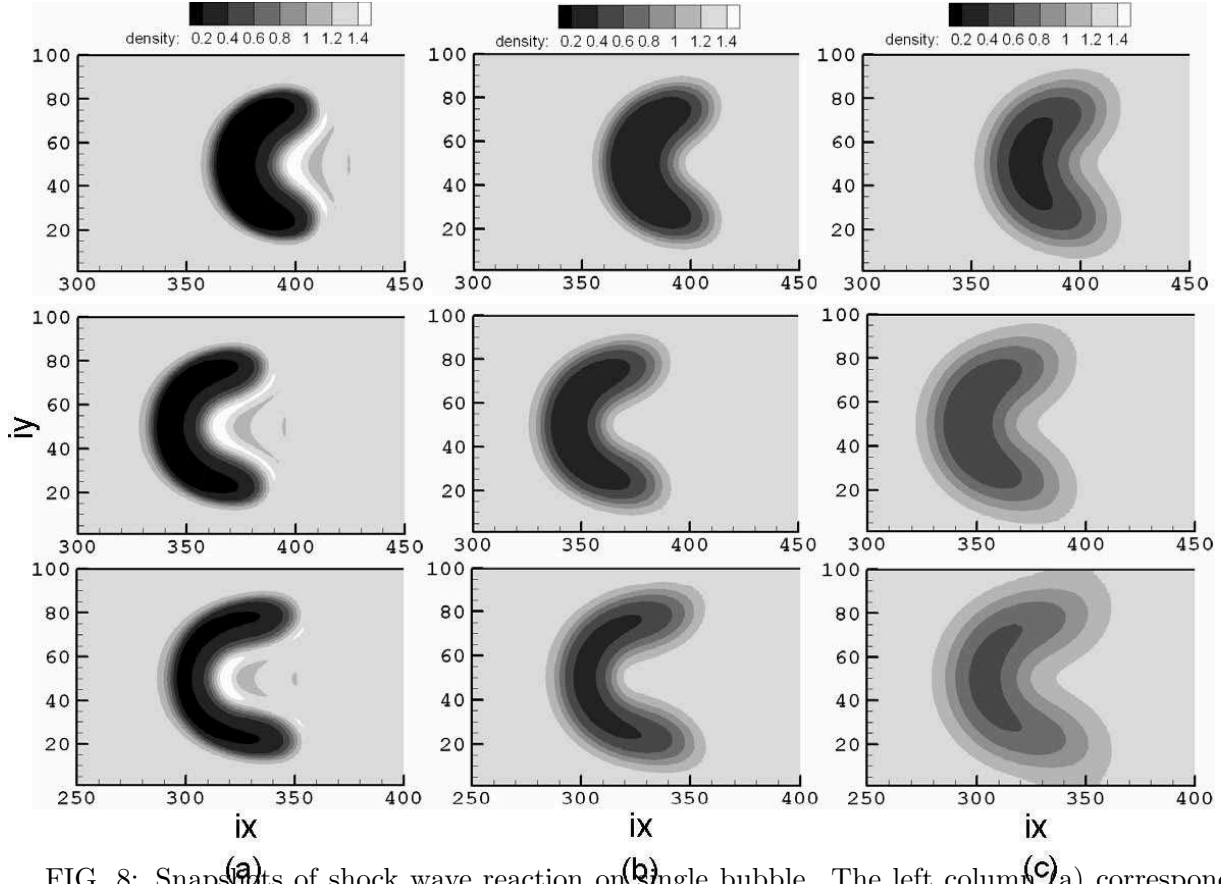}
\end{center}
\caption{Snapshots of shock wave reaction on single bubble. The left
column (a) corresponds to $s_{5}=s_{6}=10^{3}, s_{7}=s_{8}=10^{5}$,
the middle column (b) corresponds to $s_{5}=s_{6}=10^{3},
s_{7}=s_{8}=10^{4}$, and the right column (c) corresponds to
$s_{5}=s_{6}=10^{3}, s_{7}=s_{8}=10^{3}$. From black to white the
grey level corresponds to the increase of density.}
\end{figure}

The investigation of the interaction of a planar shock with an
isolated gas bubble is of special significance in the study of RM
instability, because the interface of gas bubble has typical three
dimensional characteristic and large initial distortion. It helps to
understand the mechanism of RM instability process. The problem we
simulated is as follows: A planar shock wave with the Mach number
$1.22$ ($D=1.725$), traveling from the right side, impinges on a
cylindrical bubble. The initial macroscopic quantities are as
follows:
\begin{equation}
\left( \rho ,u_{1},u_{2},p\right) \mid _{x,y,0}=\left\{
\begin{array}{cc}
\left( 1,0,0,1\right) , & \mathbf{pre-shock}\text{\textbf{,}} \\
\left( 1.28,-0.3774,0,1.6512\right) , & \mathbf{post-shock}\text{\textbf{,}}
\\
\left( 0.1358,0,0,1\right) , & \mathbf{bubble}\text{\textbf{,}}%
\end{array}%
\right.  \label{bubble1}
\end{equation}%
The domain of computation is a rectangle $Nx\times Ny=600\times
100$, $Nx$ and $Ny$ are the numbers of lattice node in the $x$- and
$y$- directions.
Initially, the bubble is at the position (450,50), the post-shock domain is $%
\left[501,600\right]\times \left[0,100\right]$. In the simulations,
the right side adopts the initial values of post-shock flow, the
extrapolation technique is applied at the left boundary, and
reflection conditions are imposed on the other two surfaces.

In Figures 7 and 8, we show some simulation results with different
configurations. The abscissa is for $ix$, and the vertical axis is
for $iy$, where $ix$ and $iy$ are the indexes of lattice node in the
$x$- and $y$- directions. From top to bottom, the three rows show
the density contours at times $t=0.5$, $0.7$, $1$, respectively. The common parameters are $%
dx=dy=0.003$, $dt=10^{-5}$. The collision parameters in Fig. 7(a) are $%
s_{5}=s_{6}=10^{4}$, $s_{7}=s_{8}=10^{5}$, those in Fig. 7(b) are $%
s_{5}=s_{6}=10^{4}$, $s_{7}=s_{8}=10^{4}$, and those in Fig. 7(c) are $%
s_{5}=s_{6}=10^{4}$, $s_{7}=s_{8}=10^{3}$, $10^{5}$ for the others. The
collision parameters in Fig. 8(a) are $s_{5}=s_{6}=10^{3}$, $%
s_{7}=s_{8}=10^{5}$, those in Fig. 8(b) are $s_{5}=s_{6}=10^{3}$, $%
s_{7}=s_{8}=10^{4}$, and those in Fig. 8(c) are $s_{5}=s_{6}=10^{3}$, $%
s_{7}=s_{8}=10^{3}$, $10^{5}$ for the others. Fig. 7 and Fig. 8 show
that when the viscosity is constant, the small thermal conductivity
is beneficial to the development of RM instability. Comparing Fig. 7
with Fig. 8, we find when the thermal conductivity is constant, the
small viscosity is beneficial to the development of RM instability.
The thermal conductivity and viscosity have inhibition effects on
the development of RM instability. Both the unmodified model and
the modified model get the same results.

\section{Conclusions}

We propose a MRT Lattice Boltzmann model which works not only for
the shocked compressible fluids but also for the unshocked
compressible fluids. In the new model, a key step is the
modification of the collision operators of energy flux so that
viscous coefficient in momentum equation and that in energy equation
are consistent no matter if the system is shocked or not. The
unnecessity of the modification for systems under strong shock is
analyzed. The new model is validated by some well-known benchmark
tests, including (i) thermal Couette flow, (ii) Riemann problem,
(iii) Richtmyer-Meshkov instability. The first system is unshocked
and the latter two are shocked. In all the three systems, the
Prandtl numbers effects are checked. Satisfying agreements are
obtained between the new model results and analytical ones or other
numerical results. Our previous models\cite{chenepl,chenpla} can be
revised in the same way to simulate unshocked compressible flows.

\section*{Acknowledgements}

This work is supported by the Science Foundations of LCP and CAEP [under
Grant Nos. 2009A0102005, 2009B0101012], National Basic Research Program of
China [under Grant No. 2007CB815105], National Natural Science Foundation of
China[under Grant Nos. 11071024, 11075021 and 11074300], the Fundamental
Research Funds for the Central Universities[under Grant No. 2010YS03].


\end{document}